\documentclass[a4paper,11pt]{article}
\pdfoutput=1 

\usepackage{jheppub} 
\usepackage{bbm,bm,graphicx,mathtools,color,hyperref,slashed}
\usepackage{amssymb,amsmath}

\usepackage[all]{xy}

\usepackage{spectralsequences} %package for spectral sequence (newest version of tex compiler is required)
\let\oldmathbb\mathbb
\protected\def\mathbb{\oldmathbb} %convert the fragile macro into the robust one

\graphicspath{{./Figures/}}

\newcommand{\diff}{\mathrm{d}}
\newcommand{\p}{\partial}
\newcommand{\ve}{\varepsilon}

\newcommand{\be}{\begin{equation}}      
\newcommand{\ee}{\end{equation}}      
\newcommand{\bea}{\begin{eqnarray}}      
\newcommand{\eea}{\end{eqnarray}}
\newcommand{\tr}{\mathrm{tr}}

\newcommand{\im}{\mathrm{i}}

\newcommand{\rme}{\mathrm{e}}

\newcommand{\spin}{\mathrm{spin}}
\newcommand{\pt}{\mathrm{pt}}
\newcommand{\CP}{\mathbb{C}P}
\newcommand*{\bZ}{\mathbb{Z}}

\usepackage{braket}
\newcommand{\s}{\sigma}
\newcommand{\g}{\gamma}

\newcommand{\C}{\mathbb{C}}
\newcommand{\Z}{\mathbb{Z}}

\newcommand{\bx}{{\bm{x}}}
\newcommand{\bk}{{\bm{k}}}

\def\widebar{\accentset{{\cc@style\underline{\mskip10mu}}}} %widebar
\def\wideubar{\underaccent{{\cc@style\underline{\mskip10mu}}}} %wideubar

\preprint{YITP-21-17}

\title{Topological terms of (2+1)d flag-manifold sigma models}

\author[1]{Ryohei Kobayashi,} 

\author[2]{Yasunori Lee,}

\author[3]{Ken Shiozaki,}

\author[3]{Yuya Tanizaki}

\affiliation[1]{
 Institute for Solid State Physics, University of Tokyo, Kashiwa, Chiba 277-8581, Japan
}

\affiliation[2]{
	\,Kavli Institute for the Physics and Mathematics of the Universe (WPI),\\
	\,University of Tokyo, Kashiwa, Chiba 277-8583, Japan
}
\affiliation[3]{Yukawa Institute for Theoretical Physics,  Kyoto University, Kyoto 606-8502, Japan}

\emailAdd{kobayashir061@gmail.com}
\emailAdd{yasunori.lee@ipmu.jp}
\emailAdd{ken.shiozaki@yukawa.kyoto-u.ac.jp}
\emailAdd{yuya.tanizaki@yukawa.kyoto-u.ac.jp}

\abstract{
We examine topological terms of $(2+1)$d sigma models and their consequences in the light of classifications of invertible quantum field theories utilizing bordism groups.
In particular, we study the possible topological terms for the $U(N)/U(1)^N$ flag-manifold sigma model in detail. 
We argue that the Hopf-like term is absent, contrary to the expectation from a nontrivial homotopy group $\pi_3(U(N)/U(1)^N)=\mathbb{Z}$, and thus skyrmions cannot become anyons with arbitrary statistics. 
Instead, we find that there exist ${N(N-1)\over 2}-1$ types of Chern-Simons terms, some of which can turn skyrmions into fermions, and we write down explicit forms of effective Lagrangians.
}

\begin{document}
\maketitle
%--------------------------------------------------------------------------------------

\section{Introduction and Summary}\label{sec:introduction}

Nonlinear sigma models appear as low-energy effective descriptions in broad areas of modern physics, ranging from high-energy to condensed-matter physics.
For example, they have allowed systematic analyses of strong-interaction phenomena~\cite{Coleman:1969sm, Callan:1969sn} even before the establishment of quantum chromodynamics (QCD),
while they have also provided useful tools to study the long-range physics of anti-ferromagnetic spin systems~\cite{Haldane:1983ru}.

In nonlinear sigma models, Lagrangian consists not only of the usual kinetic term but also of topological terms. 
Schematically, the kinetic term can be written as 
\be
\mathcal{L}_{\mathrm{kin}}=\int g_{ab}(\sigma)\,\diff \sigma^a \wedge \star \diff \sigma^b, 
\label{eq:kinetic_term}
\ee
where $\sigma:M_d\to X$ is a map from the spacetime $M_d$ to the target space $X$, and $g_{ab}(\sigma)$ is the metric of $X$, which defines the effective coupling constant. 
In practical use, we can add higher-derivative corrections to this Lagrangian, but (\ref{eq:kinetic_term}) is often sufficient for our purpose.

Let us consider what kind of topological terms can be added to this Lagrangian. 
According to the standard lore, they are classified by the homotopy of the target space $\pi_*(X)$~\cite{Coleman198802}. 
In the path integral, we are interested in the configuration with the finite Boltzmann weight $\exp(-\mathcal{L}_{\mathrm{kin}}[\sigma])<\infty$. 
Taking the infinite-volume limit $M_d\to \mathbb{R}^d$, in order to have a finite action, the sigma-model field must be constant as it approaches the infinities, 
\be
\lim_{|x|\to \infty}\sigma(x)=\sigma_*\in X.
\ee
As a result, we can identify the infinities of the spacetime with a single point, and we may regard $\sigma:S^d\to X$.
By definition, the topological nature of such maps is classified by $\pi_d(X)$.
Based on this kind of argument, it has been considered that there is a theta term corresponding to nontrivial $\pi_d(X)$,
and also that there is a Wess-Zumino term~\cite{Wess:1971yu,Witten:1983tw} when $\pi_{d+1}(X)=\bZ$ while $\pi_{d}(X)=0$, as it allows us to extend the sigma-model configuration on $S^d$ to a disk $D^{d+1}$ to define the well-defined phase factor mod $2\pi$. 

On the other hand, recent developments have led to more refined understanding of topological terms. 
Assuming quantum field theories (QFTs) can be defined on any spacetime manifolds (with given structures),
the principle of locality requires their partition functions to be consistent under cutting and gluing of spacetime manifolds.
Along this line together with the unitarity, (exponentiated) topological terms should be regarded as partition functions of the so-called invertible QFTs~\cite{Freed:2004yc}.
As a result, for the case at hand where the sigma-model target space is $X$,
%and the spacetime manifolds are spin,
topological terms are classified up to continuous ones by a group $\mathrm{Inv}_{\text{}}^{d}(X)$~\cite{Freed:2006mx, Freed:2016rqq}, 
which sits in the middle of the short exact sequence
\begin{equation}
	0
	\longrightarrow
	\mathrm{Ext}(\Omega^{\text{}}_{d}(X),\bZ)
	\longrightarrow
	\mathrm{Inv}_{\text{}}^{d}(X)
	\longrightarrow
	\mathrm{Hom}(\Omega^{\text{}}_{d+1}(X),\bZ)
	\longrightarrow
	0.
\end{equation}
In addition, ordinary theta angles invisible in $\mathrm{Inv}_{\text{}}^d(X)$ are classified by $\mathrm{Hom}(\Omega_d^{\text{}}(X),U(1))$ along with discrete theta angles~\cite{Yonekura:2018ufj}.
An important point is that these classifications use bordism groups $\Omega_\ast^{\text{}}(X)$\footnote{
	In this paper, we consider \textit{spin} bordism groups $\Omega^{\spin}_{\ast}(X)$ since the underlying microscopic theories are assumed to contain fermions.
} rather than homotopy groups $\pi_\ast(X)$,
and correspondingly the traditional arguments relying on homotopy groups might miss some subtleties.
This indeed turns out to be the case,
and one of the examples regarding the Hopf term in 3d $\CP^1$ sigma model was recently pointed out by Freed, Komargodski, and Seiberg~\cite{Freed:2017rlk}.
In this paper, we will further explore 3d sigma models with target spaces $\CP^{N-1}$ and flag manifolds $U(N)/U(1)^{N}$,
both of which can be viewed as a generalization of the simplest $\CP^1$ case.\footnote{
	Generalized (co)homologies including bordism have been also effectively used to analyze the nonlinear sigma model descriptions of 4d QCD-like theories~\cite{Freed:2006mx, Lee:2020ojw, Yonekura:2020upo}.
}\footnote{
	Analyses on the sigma-model dynamics from the bordism perspective have also been carried out in~\cite{Thorngren:2017vzn}.
}

In passing, we mention another motivation to consider the flag-manifold sigma models.
In $(1+1)$d, the sigma models with flag-manifold target space have appeared as effective descriptions of the $SU(N)$-analogue of antiferromagnetic Heisenberg-like chains~\cite{Bykov:2011ai, Bykov:2012am, Bykov:2015pka, Lajko:2017wif}, and are known to show a rich phase diagram thanks to various theta terms~\cite{Tanizaki:2018xto, Ohmori:2018qza, Hongo:2018rpy, Wamer:2020inf} (see \cite{Affleck:2021ypq} for a review).
Similarly, the $(2+1)$d flag-manifold sigma models describe the $2$d quantum $SU(3)$ Heisenberg model on several lattices~\cite{PhysRevB.88.184430, PhysRevB.95.184427}, and some of them are equivalent to the Tsunetsugu-Arikawa point~\cite{tsunetsugu2006} of the $SU(2)$ spin-$1$ model. 
It is expected that the $SU(N)$ anti-ferromagnetic Heisenberg systems may be realized with the ultracold atoms using alkaline earth metals~\cite{ PhysRevLett.92.170403, 1367-2630-11-10-103033, gorshkov2010two, taie20126, Zhang1467, 0034-4885-77-12-124401, CAPPONI201650}, and thus theoretical predictions on these systems may be able to be verified by such experiments. 
Therefore, it is an interesting question to ask what kind of topological terms can enrich those quantum spin systems.\footnote{Furthermore, $4$d flag-manifold sigma models have appeared in the study of $SU(N)$ gauge theories~\cite{Faddeev:1998yz, Amari:2018gbq}, and our analysis can be used to identify possible theta terms in those models. } 

Let us summarize our result on the $U(N)/U(1)^{N}$ flag-manifold sigma model, which shall be discussed in detail in Section~\ref{sec:flag}. 
Although the homotopy argument suggests the presence of Hopf-like term because of the nontrivial homotopy $\pi_3(U(N)/U(1)^{N})=\mathbb{Z}$, we conclude that such a term does not exist since
\be
\Omega^{\mathrm{spin}}_3(U(N)/U(1)^{N})=0
\ee
for $N\ge 3$, assuming that the spacetime manifold is equipped with spin structure. 
We also find that there are ${N(N-1)\over 2}-1$ types of Chern-Simons-like terms according to the bordism computation, and we try to write down their explicit forms using the gauged sigma-model description. 
In such a description, we introduce $U(1)$ gauge fields $a_1,\ldots, a_{N-1}$, which indeed allows us to write down the Chern-Simons coupling, but there still exists a puzzle. 
When naively writing down possible Chern-Simons term, we get 
\be
\sum_{I=1}^{N-1} {k_I \over 4\pi}a_I\diff a_I + \sum_{I<J}{ k_{IJ}\over 2\pi} a_I \diff a_J, 
\ee
and there are ${N (N-1)\over 2}$ of them, which overcounts the number of independent Chern-Simons couplings compared to the bordism computation. 
It turns out that there is a nontrivial relation, which allows us to eliminate one of the off-diagonal Chern-Simons couplings, and we find there exist globally well-defined $3$-forms, which can be added to the effective Lagrangian, as a byproduct. 
For example for $N=3$, the topological terms of the effective Lagrangian are given as
\be
{k_1\over 4\pi}a_1\diff a_1+{k_2\over 4\pi}a_2\diff a_2+{h\over 2\pi}a_{12}a_{23}a_{31}, 
\ee
where the first two terms are diagonal Chern-Simons terms, and the last one is a non-quantized topological coupling with $a_{ij}$ defined as \eqref{aij}, while the off-diagonal Chern-Simons term is absent. 
In particular, when the diagonal Chern-Simons terms have odd levels, skyrmions are transformed as fermions under spacetime rotations. 

The rest of the paper is organized as follows.
In Section~\ref{sec:CP1}, we review the case of 3d $\CP^1$ sigma model.
We first directly compute bordism groups to identify the topological (``Hopf'') term,
and then discuss its consequence for the statistics of skyrmions from the physics point of view, using gauged realization of the model.
In Sections~\ref{sec:CPN} and \ref{sec:flag}, we further go on to the case of 3d sigma model with target spaces $\CP^{N-1}$ and flag manifolds $U(N)/U(1)^N$, respectively.
Appendix~\ref{app:CSconstraints} serves as a supplement involving some subtleties arising in the flag manifold case.
In Appendix~\ref{app:LSSS}, various relevant cohomology groups are computed via Leray-Serre spectral sequence (LSSS).
In Appendix~\ref{sec:fermion_sigma}, we review the realization of the Hopf term of the $\C P^1$ model as a partition function of a fermionic invertible phase~\cite{Abanov:1999qz}.

\section{Review on the topological term of 3d $\mathbb{C}P^1$ model}\label{sec:CP1}

$3$d $\mathbb{C}P^1$ sigma model gives a low-energy description of the $2$d anti-ferromagnetic quantum spin systems. 
Because of $\pi_3(\mathbb{C}P^1)=\mathbb{Z}$ which represents the Hopf fibration, it has been suspected for more than three decades that the model accommodates a topological theta term, which can be schematically written as 
\be
\im\theta\, \mathrm{Hopf}[n]. 
\ee
The paper by Wilczek and Zee~\cite{Wilczek:1983cy} has shown that the (magnetic) skyrmion acquires the fractional spin $\theta/2\pi$ and becomes an anyon due to this Hopf term.
The Hopf term does not have a local expression in the original spin fields, but can be written as a Chern-Simons form using the $U(1)$ gauge field in the gauged sigma-model description~\cite{Wu:1984kd}.
%as shown by Wu and Zee~\cite{Wu:1984kd}, and they also confirmed the fractional spin induced by the Hopf term. 

Meanwhile, starting from the $(2+1)$d $SU(2)$ anti-ferromagnetic quantum Heisenberg  spin system, it is known that the Hopf term does not appear in the low-energy theory~\cite{Haldane:1988zz}.
Instead, the discrete value of the theta term $\theta=\pi$ arises when the spin variable is coupled to fermions by Yukawa interaction, after integrating out fermions~\cite{Abanov:1999qz}.

In a recent paper~\cite{Freed:2017rlk}, it has been elucidated that only these values, $\theta=0$ and $\theta=\pi$, are consistent as local and unitary QFTs. 
Indeed, $k=\theta/\pi$ behaves as the level of $U(1)$ spin Chern-Simons term, and the magnetic skyrmion becomes fermion if $k$ is odd. 

In the following of this section, we will give a review on these facts, as it is a basic ingredient in order to extend these results for general flag-manifold sigma models. 

\subsection{Formal and explicit descriptions of topological terms of $\mathbb{C}P^1$}

Here, we first give a formal description on the possible topological terms in the $\mathbb{C}P^1$ sigma model, and then write down the explicit forms for the topological term. 
Unless explicitly stated, we always assume that the spacetime manifolds are equipped with spin structure.

As briefly explained in Sec.~\ref{sec:introduction}, while topological terms have been conventionally classified using homotopy,
the correct framework to be adopted is bordism rather than homotopy 
when we require the generalized locality (i.e. consistency under cutting and gluing rules) for QFTs, 
so that they can be defined on any manifolds beyond spheres.
Accordingly, topological terms of the $3$d $\mathbb{C}P^1$ sigma model should be captured by spin bordism groups $\Omega^{\mathrm{spin}}_* (\mathbb{C}P^1)$
from this modern point of view\footnote{
	Strictly speaking, the first two statements are established as theorems,
	while the last statement is still better to be regarded as a conjecture.
	Note, however, that since it can be viewed as a formal way to present the Stora-Zumino descent equation~\cite{Stora:1983ct, Zumino:1983ew}, the authors think it quite reasonable.
}:
\begin{itemize}
	\item $U(1)$ factors in $\mathrm{Hom}(\Omega^{\text{spin}}_{3}(\CP^1),U(1))$ 
% 	or equivalently $\mathrm{Free}(\Omega^{\text{spin}}_{3}(\CP^1))$
    classifies the ordinary theta angles~\cite{Yonekura:2018ufj},
	\item $\mathrm{Ext}(\Omega^{\text{spin}}_{3}(\CP^1),\bZ) \simeq \mathrm{Tors}(\Omega^{\text{spin}}_{3}(\CP^1))$ classifies the discrete theta angles~\cite{Freed:2006mx, Freed:2016rqq},
	\item $\mathrm{Hom}(\Omega^{\text{spin}}_{4}(\CP^1),\bZ) \simeq\mathrm{Free}(\Omega^{\mathrm{spin}}_4 (\mathbb{C}P^1))$ classifies the Chern-Simons or Wess-Zumino terms~\cite{Freed:2006mx, Freed:2016rqq}.
\end{itemize}
These groups can be computed using the Atiyah-Hirzebruch spectral sequence (AHSS)
\begin{equation}
	E^2_{p,q}
	=
	H_p(\mathbb{C}P^1 ; \Omega^{\mathrm{spin}}_q(\mathrm{pt}))
	\Longrightarrow
	\Omega^{\mathrm{spin}}_{p+q} (\mathbb{C}P^1), 
\end{equation}
where the $E^2$-page on the left hand side converges to the desired group on the right hand side.
The explicit form of the $E^2$-page is given as follows:
\begin{equation}
\begin{sseqpage}[
	homological Serre grading,
	grid = chess ,
	classes = {draw = none},
	yscale = 0.7, 
	x label = {$H_p(\mathbb{C}P^1)$},
	y label = {$\Omega^{\mathrm{spin}}_q(\mathrm{pt})$}
]
	\foreach \x in {0,2}{
		\class["\mathbb{Z}"](\x,0)
		\class["\mathbb{Z}_2"](\x,1)
		\class["\mathbb{Z}_2"](\x,2)
		\class["\mathbb{Z}"](\x,4)
	}
	\d[dashed]2(2,1)
	\class[""](3,0)
	\class[""](4,0)
\end{sseqpage}
\end{equation}
It is known that, if the space $X$ is connected, bordism groups split as
\begin{equation}
	\Omega_\ast^{\text{spin}}(X) = \Omega_\ast^{\text{spin}}(\pt) \oplus \widetilde \Omega_\ast^{\text{spin}}(X),
\end{equation}
and therefore differentials going into the $p=0$ column (e.g.~the dashed arrow) are all trivial.
As a result, $E^2_{2,1}=H_2(\mathbb{C}P^1; \Omega^{\mathrm{spin}}_1(\mathrm{pt}))= \mathbb{Z}_2$ survives to the $E^\infty$-page, and one has
\begin{equation}
	\renewcommand{\arraystretch}{1.3}
	\begin{array}{ccc}
		\mathrm{Free}(\Omega_3^{\text{spin}}(\CP^1)) & = & 0,\\
		\mathrm{Tors}(\Omega_3^{\text{spin}}(\CP^1)) & = & \bZ_2,\\
		\mathrm{Free}(\Omega_4^{\text{spin}}(\CP^1)) & = & \bZ.
	\end{array}
\end{equation}
The free part of $\Omega_4^{\mathrm{spin}}(\mathbb{C}P^1)$ comes from $\Omega_4^{\mathrm{spin}}(\pt)$ describing the gravitational Chern-Simons term, which is not of our interest; our interest will only be in $\widetilde\Omega_4^{\text{spin}}(\CP^1)$.
The Hopf term with continuous $\theta\in \mathbb{R}/2\pi \bZ$ parameter is in fact absent as the free part of $\Omega_3^{\mathrm{spin}}(\mathbb{C}P^1)$ is trivial.
We identify the torsion part of $\Omega_3^{\mathrm{spin}}(\mathbb{C}P^1)$ as the ``Hopf'' term of 3d $\mathbb{C}P^1$ model, but its theta angle should be discretized as it is $\mathbb{Z}_2$-valued. 
% This is exactly what we have found in the (lengthy) discussion in previous sections.

Let us move on to explicit computations in order to write down the discrete theta term.
%for $\mathrm{Tors}(\Omega^{\mathrm{spin}}_3(\mathbb{C}P^1))$. 
We can realize the $\mathbb{C}P^1$ sigma model as a $U(1)$ gauge theory according to the fibration $U(1)\to SU(2)\to \mathbb{C}P^1$, that is, we introduce the $SU(2)$-valued scalar field $U(x)$ and the $U(1)$ gauge field $a=a_\mu \diff x^\mu$. 
We note that $SU(2)$ matrix $U$ can be written as 
\be
U=\begin{pmatrix} 
z_1 & -z_2^*\\
z_2 & z_1^*
\end{pmatrix}, 
\ee
with $|z_1|^2+|z_2|^2=1$, and thus the $SU(2)$ scalar field is equivalent to the two-component complex scalar field $z=(z_{\alpha})_{\alpha=1,2}$ with a unit norm. 
Assigning $U(1)$ gauge charge $1$ to this scalar $z$, the kinetic term in the Lagrangian is given by 
\be
{1\over g^2}|(\p_\mu+\im a_\mu)z|^2={1\over g^2}\left(|\p_\mu z|^2+\im a_\mu (-z^\dagger \p_\mu z+(\p_\mu z^\dagger )z)+a_\mu^2\right). 
\ee
Since this is quadratic in $a_\mu$, the equation of motion can be solved as 
\be
a=\im z^\dagger \diff z. 
\label{eq:gauge_field_CP1}
\ee
With this $U(1)$ gauge field, we define the Hopf term as the Chern-Simons term, following the idea of \cite{Wu:1984kd}. 
That is, we extend the spacetime manifold $M_3$ to the $4$d spin manifold $W_4$ with $\p W_4=M_3$ and define 
\be
\exp\Bigl(\im (\pi k) \mathrm{Hopf}[M_3,(z,a)]\Bigr):=\exp\left(\im {k\over 4\pi}\int_{W_4}\diff \tilde{a}\wedge \diff \tilde{a}\right),
\label{eq:definition_Hopf}
\ee
where $\tilde{a}$ is the $4$d extension of $a$ \textit{as $U(1)$ gauge fields}. 

We shall give concrete computations to show that this gives the correct discrete theta term in the next subsection. 
Before that, let us make several remarks, which would be confusing at first sight.

In the definition (\ref{eq:definition_Hopf}), we only extend the $U(1)$ gauge field $a$, while the $\mathbb{C}P^1$ field needs not be extended to $W_4$, and therefore (\ref{eq:gauge_field_CP1}) holds only on the boundary $M_3$ but not on the $4$d bulk.\footnote{In the work of Abanov~\cite{Abanov:2000ea}, the similar idea has been used to find the Hopf term by integrating out the fermion when the spin variable $\bm{n}=z^\dagger \bm{\sigma} z$ couples to it. Instead of directly evaluating the fermion determinant, they consider an infinitesimal variation of the spin variable, and try to integrate it to obtain the result. Within the $\mathbb{C}P^1$ model, however, the Hopf term is a total derivative~\cite{Wu:1984kd}, and the infinitesimal variation identically vanishes. They circumvented this issue by embedding $\mathbb{C}P^1\subset \mathbb{C}P^{N-1}$ with $N>2$. As $\mathrm{B}U(1)=\mathbb{C}P^{\infty}$, our prescription (\ref{eq:definition_Hopf}) can be formally regarded as the $N\to \infty$ limit of it.} 
We note that $\Omega^{\mathrm{spin}}_{3}(\mathbb{C}P^1)= \mathbb{Z}_2$ shows that the $\mathbb{C}P^1$ field may not have a $4$d extension when the spin structure is taken into account: 
Only after taking the double copy, $M_3\sqcup M_3$, we can take the $4$d bounding manifold $M'_4$ such that the extension of the $\mathbb{C}P^1$ field is possible. 

Next, let us prove that the Hopf term defined in \eqref{eq:definition_Hopf} is a bordism invariant of $3$d spin manifolds equipped with a map $\sigma=(z,a)$ to $\CP^1$.
To see this, suppose the pair $(M_3, \sigma)$ is bordant to $(M_3', \sigma')$, i.e., there exists a 4d spin manifold $Y_4$ and a map $\widetilde{\sigma}: Y_4\to\CP^1$, such that $\partial Y_4=M_3\sqcup \overline{M'_3}$ and 
\be
\widetilde{\sigma}\bigr|_{M_3}=\sigma,\quad 
\widetilde{\sigma}\bigr|_{\overline{M'_3}}=\sigma'. 
\ee
Then we have
\begin{align}
    \frac{\exp\Bigl({\im (\pi k)\mathrm{Hopf}[M_3,\sigma]}\Bigr)}{\exp\Bigl({\im (\pi k)\mathrm{Hopf}[M'_3,\sigma']}\Bigr)}=\exp\left(\im {k\over 4\pi}\int_{Y_4} \diff \tilde{a}\wedge \diff \tilde{a}\right)=1.
\end{align}
Here, unlike in the definition (\ref{eq:definition_Hopf}), the $4$d extension to $Y_4$ is taken as a $\mathbb{C}P^1$ field, and thus the $U(1)$ gauge field is subject to the constraint (\ref{eq:gauge_field_CP1}), $\tilde{a}=\im \tilde{z}^\dagger \diff \tilde{z}$. As a result, $\diff \tilde{a}\wedge \diff \tilde{a}=0$ identically. 
% We note that the $U(1)$ gauge field $a$ is defined via the Maurer-Cartan form of the $SU(2)$ scalar field~\eqref{eq:gauge_field_CP1}, and the 4d theta term must vanish $f\wedge f=0$ since $SU(2)$ is three dimensional. 
Hence, we can see that
\begin{align}
    \exp\Bigl({\im (\pi k)\mathrm{Hopf}[M_3,\sigma]}\Bigr)=\exp\Bigl({\im (\pi k)\mathrm{Hopf}[M'_3,\sigma']}\Bigr),
\end{align}
which means that the Hopf term (\ref{eq:definition_Hopf}) gives a bordism invariant.

\subsection{Statistics of skyrmions}\label{sec:statistics_cp1}

Let us illustrate how the level $k$ of the Hopf term is quantized. 
Since the 3d spin bordism group equipped with a map to $\CP^1$ is given by $\Omega_3^{\mathrm{spin}}(\CP^1)=\mathbb{Z}_2$, the Hopf term must take a $\mathbb{Z}_2$ valued phase, $\exp\Bigl({\im \pi\, \mathrm{Hopf}[M_3,\sigma]}\Bigr)=\pm 1$. Therefore, we should identify 
\be
k\sim k+2, 
\label{eq:level_identify_cp1}
\ee
in the path-integral weight, $\exp\Bigl({\im (\pi k) \mathrm{Hopf}[M_3,\sigma]}\Bigr)$. 
This shows that we can verify the definition (\ref{eq:definition_Hopf}) of the Hopf term by constructing a $\mathbb{C}P^1$ configuration $\sigma$ on a $3$-manifold $M_3$ which gives $\exp\Bigl({\im \pi \mathrm{Hopf}[M_3,\sigma]}\Bigr)=- 1$ for $k=1$.  

We can construct such a configuration by taking the spacetime $M_3$ to be $S^2\times S^1$, introducing a magnetic skyrmion on the space $S^2$, and then performing the adiabatic $2\pi$ rotation along the imaginary time $S^1$. This is exactly the way to measure the statistics of the skyrmion in a semi-classical manner~\cite{Wilczek:1983cy, Wu:1984kd}. 
A skyrmion can be put on $S^2$ by setting 
\be
z(\theta,\phi)=\begin{pmatrix}
\cos{\theta\over 2}\\
\rme^{\im\phi}\sin{\theta\over 2}
\end{pmatrix},
\ee
where $(\theta,\phi)$ is the spherical coordinate of $S^2$ with $0\le \theta \le \pi$ and $0\le \phi <2\pi$. 
We can perform the adiabatic $2\pi$ rotation along the imaginary time $0\le \tau<2\pi$ as 
\be
z(\tau,\theta,\phi):=\rme^{\im \tau/2}\exp\left(\im{\tau\over 2}\sigma_3\right)z(\theta,\phi)
=\begin{pmatrix}
\rme^{\im \tau}\cos{\theta\over 2}\\
\rme^{\im\phi}\sin{\theta\over 2}
\end{pmatrix}.
\label{cp1soliton}
\ee
Here, $\exp\left(\im{\tau\over 2}\sigma_3\right)$ describes the adiabatic rotation, and we further multiply the overall $U(1)$ phase $\rme^{\im \tau/2}$ so that the configurations at $\tau=0$ and $\tau=2\pi$ are in the same gauge.
We then obtain the $U(1)$ gauge field $a$ on $S^2\times S^1$ from (\ref{cp1soliton}) as 
\be
a=-\cos^2{\theta\over 2}\,\diff \tau-\sin^{2}{\theta\over 2}\, \diff \phi.  
\label{u1monopole}
\ee

In order to evaluate (\ref{eq:definition_Hopf}), we need to extend (\ref{u1monopole}) from $S^2\times S^1$ to a $4$-manifold. We can take $W_4=S^2\times D^2$ as an extension, where the polar coordinate of $D^2$ is given by $(\rho,\tau)$ and the extension of $a$ is 
\be
\tilde{a}=-\cos^2{\theta\over 2}\,\rho^2 \diff \tau-\sin^{2}{\theta\over 2}\, \diff \phi.
\ee
We can readily check that 
\be
{1\over 4\pi}\int_{S^2\times D^2}\diff \tilde{a}\wedge \diff \tilde{a}=\pi, 
\ee
which confirms that $\exp\left(\im \pi \mathrm{Hopf}[S^2\times S^1, (z,a)]\right)=-1$. 
This means that the skyrmion becomes fermionic due to the Hopf term when we set $k=1$. 

One should be able to confirm the statistics of the skyrmion also by checking the statistics of its creation/annihilation operator. 
Since a single skyrmion in the space $S^2$ introduces the configuration of the $U(1)$ gauge field~\eqref{u1monopole}, we identify its creation/annihilation operator as the monopole operator $m(x)$ for the $U(1)$ gauge field. 
The monopole operator $m(x)$ is defined as a defect operator. Roughly speaking, we remove a point $x$ from the spacetime $3$-manifold $M_3$, and require the $U(1)$ gauge field $a$ to behave as 
\be
\int_{S^2_x} {1\over 2\pi}\diff a =\pm 1, 
\ee
where $S^2_x$ is a sufficiently small two-sphere surrounding the removed point $x$. When the monopole operator $m(x)$ is inserted, we assert that the path integral is performed over the gauge fields with this boundary condition. 

We note that the monopole operator is not $U(1)$ gauge invariant when the Chern-Simons term exists~\cite{Pisarski:1986gr, Affleck:1989qf}: Under the gauge transformation, $a\mapsto a+\diff \varepsilon$ and $z\mapsto \rme^{-\im \ve} z$, 
\be
m(x)\mapsto \rme^{\im \,k\, \ve(x)}m(x). 
\ee
As a result, the gauge-invariant monopole operator should be defined as 
\be
m(x) z_{\alpha_1}(x)\cdots z_{\alpha_k}(x). 
\ee
The quantum number of this operator should be identical with that of the corresponding skyrimon. 
In order to find the spin of the gauge-invariant monopole operator, note that the spherical coordinate dependence of the scalar field $z$ is described by the monopole harmonics, $Y_{q,\ell,m}$, near the monopole singularity~\cite{Borokhov:2002ib}. 
For the minimal monopole charge, $q=1/2$, and the allowed values of $\ell$ are $\ell\in |q|+\mathbb{Z}_{\ge 0}$~\cite{Wu:1976ge, Wu:1977qk}. 
Therefore, the spin of the monopole operator is given by $k/2$ mod $\mathbb{Z}$, which is consistent with the above computations.

%------------------------------------------------------------------------
\section{Topological term of 3d $\mathbb{C}P^{N-1}$ sigma model with $N\ge 3$}\label{sec:CPN}

Let us move on to the $3$d $\mathbb{C}P^{N-1}$ nonlinear sigma model, and consider its topological terms. 
Even though the expressions for the topological terms of $\mathbb{C}P^1$ and $\mathbb{C}P^{N-1}$ are superficially similar, we shall emphasize the difference between them in this section.
This understanding turns out to be useful to find the correct description of topological terms for the flag-manifold model in the next section. 

\subsection{Conflict between homotopy argument and the gauged-sigma model}

The $\mathbb{C}P^{N-1}$ sigma model can be realized using a $\mathbb{C}^N$-valued scalar field $z=(z_{\alpha})_{\alpha=1,\ldots, N}$ with a unit norm, with a $U(1)$ gauge field $a$. 
This corresponds to the fibration
\be
S^1\to S^{2N-1} \to \mathbb{C}P^{N-1}, 
\label{eq:CPn-1_fibration}
\ee
which is identical with the Hopf fibration for $N=2$. 
In the following, we consider the case with $N\ge 3$. 
As a consequence of the homotopy exact sequence, for $k\ge 3$, 
\be
\pi_k(S^{2N-1})= \pi_k(\mathbb{C}P^{N-1})
\ee
as $\pi_k(S^1)=\pi_{k-1}(S^1)=0$, and we immediately find that 
\be
\pi_3(\mathbb{C}P^{N-1})= \pi_4(\mathbb{C}P^{N-1})=0.
\ee
Therefore, in the conventional view of the homotopy, we do not expect the $3$d $\mathbb{C}P^{N-1}$ model to have any topological terms, neither theta terms nor Wess-Zumino terms.

However, if we regard the $\mathbb{C}P^{N-1}$ model as an example of $U(1)$ gauge-Higgs systems, we can write down a Chern-Simons coupling for the $U(1)$ gauge field, 
\be
{\im k\over 4\pi}\int_{W_4} \diff \tilde{a} \wedge \diff \tilde{a},
\label{eq:CS_cpn-1}
\ee
where $W_4$ is a four-dimensional extension of the spacetime $3$-manifold along with the spin structure. 
We would like to resolve the discrepancy between these two arguments, and also find the physical consequence of the topological terms.

\subsection{Spin bordism of $\mathbb{C}P^{N-1}$ and the topological term}

To find possible topological terms,
we again consider AHSS
\be
E^2_{p,q} = H_p(\mathbb{C}P^{N-1}; \Omega^{\spin}_{q}(\pt))\Longrightarrow \Omega^{\spin}_{p+q}(\mathbb{C}P^{N-1}). 
\ee 
The $E^2$ page is given by the following figure:
\be
\begin{sseqpage}[homological Serre grading, grid = chess , classes = {draw = none}, yscale = 0.7, 
	 x label = {$H_p(\mathbb{C}P^{N-1})$}, y label = {$\Omega^{\mathrm{spin}}_q(\mathrm{pt})$}
	]
\foreach \x in {0,2,4,6}{
	\class["\mathbb{Z}"](\x,0)
	\class["\mathbb{Z}_2"](\x,1)
	\class["\mathbb{Z}_2"](\x,2)
	\class["\mathbb{Z}"](\x,4)
}
\d2(4,0)
\d2(4,1)
\end{sseqpage}
\ee
where the homology of $\mathbb{C}P^{N-1}$ is
\be
H_*(\mathbb{C}P^{N-1})= \left\{
\begin{array}{ccc}
\mathbb{Z} &\quad *=0,2,\ldots, 2(N-1), \\
0 &\quad \mathrm{else}, 
\end{array}
\right.
\label{eq:homology_CPn-1}
\ee
(see Appendix~\ref{sec:LSSS_CPn-1} to confirm this result from the Leray-Serre spectral sequence (LSSS)).
Important difference from the $N=2$ case is that $H_4(\mathbb{C}P^{N-1})=\mathbb{Z}$ for $N\ge 3$ and
this gives rise to a possibly-nontrivial differential $d^2: E^2_{4,0}\to E^2_{2,1}$. 
This map is known to be given by a dual of Steenrod square composed with mod $2$ reduction~\cite{Teichner:MR1214960},
and here the generator $x\in H^2(\mathbb{C}P^{N-1}; \mathbb{Z}_2)$
is mapped to $\mathrm{Sq}^2(x)=x\cup x\in H^4(\mathbb{C}P^{N-1}; \mathbb{Z}_2)$. Correspondingly $d^2$ is nontrivial, and thus $E^2_{2,1}=\bZ_2$ is eliminated in the $E^{\infty}$-page, leading to
\be
\Omega^{\mathrm{spin}}_3(\mathbb{C}P^{N-1})=0\qquad (N\ge 3). 
\ee
Therefore, (not only the ordinary ones but also) the discrete theta angles do not exist in the $3$d $\mathbb{C}P^{N-1}$ sigma model for $N\ge 3$.
Instead, we have 
\be
\widetilde\Omega^{\spin}_4(\mathbb{C}P^{N-1})= 2\mathbb{Z}
\label{eq:CPn-1_spin_4th_bordism}
\ee
which is nothing but the Chern-Simons coupling (\ref{eq:CS_cpn-1}) in our description.
Here, the map $H_4(\CP^{N-1};\bZ)=\bZ \to 2\bZ\subset \Omega^{\spin}_4(\CP^{N-1})$ is a multiplication by two,\footnote{We would like to note that this factor $2$ appears due to the spin structure of the spacetime. When we only assume the orientation structure, we need to compute the oriented bordism, instead. As (reduced) oriented bordism is exactly identical to the homology at low degrees without this extra factor, we can classify the topological terms using ordinary (co)homology in such cases~\cite{Davighi:2020vcm}. 
Use of generalized (co)homology allows us to deal with the subtle factor related to the spacetime structure. } since the latter is the kernel of $d^2:E^2_{4,0}\to E^2_{2,1}$. 
To be more explicit,\footnote{We thank the anonymous referee for suggestion of giving the following explicit explanation. } we can take $\mathbb{C}P^2\subset \mathbb{C}P^{N-1}$ as the generator of $H_4(\mathbb{C}P^{N-1},\mathbb{Z})$, which satisfies $\int_{\mathbb{C}P^2}\left({\diff a \over 2\pi}\right)^2=1$. 
On spin $4$-manifolds $X$, however, the Atiyah-Singer index theorem tells that $\int_{X} \left({\diff a\over 2\pi}\right)^2$ is always an even integer, so there is a factor-$2$ difference between the integral homology and the spin bordism of degree $4$.

Let us make several remarks. 
Although the definition~(\ref{eq:CS_cpn-1}) looks very similar to the one~(\ref{eq:definition_Hopf}) in the $\mathbb{C}P^1$ model, there are important differences. 
As $\Omega^{\mathrm{spin}}_3(\mathbb{C}P^{N-1})=0$ for $N\ge 3$, the $\mathbb{C}P^{N-1}$ configurations in $3$ dimensions are always null-bordant within the spin manifold. 
Therefore, unlike in the case of $\mathbb{C}P^1$, we can take the $4$d extension of the fields $(\tilde{z},\tilde{a})$ \textit{as $\mathbb{C}P^{N-1}$ fields} in the definition (\ref{eq:CS_cpn-1}).
A related remark is that, for $\mathbb{C}P^{N-1}$ models, we do not have any identifications on $k$, 
although the Hopf tem for the $\mathbb{C}P^1$ model has (\ref{eq:level_identify_cp1}). 
For such an identification to exist, the phase factor $\exp\left({\im \over 4\pi}\int a\diff a\right)$ needs to be quantized.
This indeed occurs for the $\mathbb{C}P^1$ model because $\diff \tilde{a}\wedge \diff \tilde{a} =0$ when $\tilde{a}$ is restricted to $\tilde{a}=\im \tilde{z}^\dagger \diff \tilde{z}$. 
However, for $\mathbb{C}P^{N-1}$ with $N\ge 3$, one can easily check that $\diff \tilde{a}\wedge \diff \tilde{a}\not =0$ even with the constraint $\tilde{a}=\im \tilde{z}^\dagger \diff \tilde{z}$, so such quantization of phases does not occur. 
This explains the fact that the topological term (\ref{eq:CS_cpn-1}) comes from $\Omega^{\mathrm{spin}}_4(\mathbb{C}P^{N-1})$, not from $\Omega^{\mathrm{spin}}_3(\mathbb{C}P^{N-1})$, in an intuitive way. 

\subsection{Statistics of skyrmions}

Let us discuss the statistics of skyrmions. We add the level-$1$ Chern-Simons coupling to $\mathbb{C}P^{N-1}$ sigma model for $N\ge 3$.

In order to discuss the quantum number of skyrmions, we can look at the statistics of corresponding monopole operators, as we have briefly explained in Sec.~\ref{sec:statistics_cp1}.
Because of the Chern-Simons coupling (\ref{eq:CS_cpn-1}), the naive monopole operator is not $U(1)$ gauge invariant, and the gauge-invariant monopole operator is given as 
\be
m(x)z_\alpha(x). 
\ee
Therefore, the skyrmion is in the defining representation of $SU(N)$ flavor symmetry.
In order to discuss its spin, we again note that the $z$ field in the vicinity of the monopole singularity should be decomposed with the monopole harmonics $Y_{1/2,\ell,m}$.
Thus, the $z$ field has a half-integer spin, and so does the monopole operator $m(x) z_{\alpha(x)}$. As a result, the corresponding skyrmion is a fermion. 

Let us confirm this result by directly evaluating the topological term (\ref{eq:CS_cpn-1}) for the adiabatic $2\pi$-rotation of a skyrmion. 
To be specific, let us set $N=3$ and put a $\mathbb{C}P^1$ skyrmion configuration in the first two components of $\mathbb{C}P^{2}$ fields:
\be
z(\theta,\phi)=\begin{pmatrix}
\cos {\theta\over 2}\\ 
\rme^{\im \phi} \sin{\theta\over 2}\\
0
\end{pmatrix}.
\ee
Here, we take the spherical coordinate of the space $S^2$,
and make $\theta$ and $\phi$ run over $[0,\pi]$ and $[0,2\pi)$ respectively. 
Then, we can realize the $2\pi$ rotation of this skyrmion by multiplying the following $SU(3)$ matrix, 
\be
\tilde{z}(\rho,\alpha,\theta,\phi)=
\begin{pmatrix}
1& 0& 0\\
0 & \rho\,\rme^{\im \alpha} & -\sqrt{1-\rho^2}\\
0 & \sqrt{1-\rho^2}& \rho\,\rme^{-\im \alpha}
\end{pmatrix}
z(\theta,\phi).
\label{eq:2pi_rotation_cpn-1}
\ee
Here, $\alpha\sim \alpha+2\pi$ is regarded as the imaginary-time circle at $\rho=1$, and it is extended to the two-dimensional disk $D^2$. 
Importantly, unlike the case of $\mathbb{C}P^1$ model, we here take a 4d extension of both the $U(1)$ gauge field and the scalar field.\footnote{
	We note that the same is true for skyrmions of $4$d $N_f$-flavor QCD; for $N_f=2$, we cannot extend the skyrmion configuration with $2\pi$ rotation into 5d, but it becomes possible if we embed it into larger flavors $N_f\geq 3$~\cite{Witten:1983tx}.
}

Now, let us evaluate the Chern-Simons term. The $U(1)$ gauge field is obtained as 
\be
\tilde{a}=\im \tilde{z}^\dagger \diff \tilde{z}
=-\sin^2 {\theta\over 2}\, \diff \phi -\sin^2{\theta\over 2}\, \rho^2 \diff \alpha. 
\ee
The phase factor acquired by the skyrmion is given by
\bea
&&\exp\left({\im\over 4\pi}\int_{S^2\times D^2}\diff \tilde{a}\wedge \diff \tilde{a}\right)\nonumber\\
&=&\exp\left({\im\over 4\pi}2\int_0^\pi \sin^2{\theta\over 2}\diff \sin^2{\theta\over 2}\int_0^{2\pi}\diff \phi\int_0^1\diff \rho^2 \int_0^{2\pi}\diff \alpha\right)\nonumber\\
&=&-1,
\eea
which shows that the skyrmion with the Chern-Simons coupling becomes a fermion.

%------------------------------------------------------------------------
\section{Topological terms of 3d flag-manifold sigma model}\label{sec:flag}
We finally consider sigma models having flag manifolds
\begin{equation}
	\dfrac{U(N)}{U(1)^N}
\end{equation}
as a target space,
which are another type of generalizations of $\CP^1=U(2)/U(1)^2$.

\subsection{Spin bordism of flag manifolds}

As before, the classification of topological terms is given in terms of $\Omega^{\spin}_{*}(U(N)/U(1)^N)$.
The AHSS we consider is 
\begin{equation}
	E^2_{p,q} = H_p(U(N)/U(1)^N; \Omega^{\spin}_{q}(\pt))\Longrightarrow \Omega^{\spin}_{p+q}(U(N)/U(1)^N)	
\end{equation}
%We note that 
%\begin{equation}
%	H_*(\mathbb{C}P^{N-1})\simeq \left\{
%	\begin{array}{ccc}
%	\mathbb{Z} &\quad *=0,2,\ldots, 2(N-1), \\
%	0 &\quad \mathrm{else}.
%	\end{array}
%	\right.
%\end{equation}
and the $E^2$ page is as follows:
\begin{equation}
	\begin{sseqpage}[
		homological Serre grading,
		grid = chess,
		classes = {draw = none},
		xscale = 2.5,
		y axis gap = 1.5cm,
		x axis extend end = 1.5cm
	]

	\class["\bZ"](0,0)
	\class["\bZ_2"](0,1)
	\class["\bZ_2"](0,2)
	\class["\bZ"](0,4)
	
	\class["\bZ^{\oplus (N-1)}"](2,0)
	\class["\bZ_2^{\oplus (N-1)}"](2,1)
	\class["\bZ_2^{\oplus (N-1)}"](2,2)
	\class["\ast"](2,4)
	
	\class["\bZ^{\oplus \left({N(N-1)\over 2}-1\right)}"](4,0)
	\class["\bZ_2^{\oplus \left({N(N-1)\over 2}-1\right)}"](4,1)
	\class["\ast"](4,2)
	\class["\ast"](4,4)
	
	\d2(4,0)
	\d2(4,1)
	\end{sseqpage}
	\label{eq:E2flag}
\end{equation}
See Appendix~\ref{sec:cohomology_flag_manifold} for the computation of integral cohomology of flag manifolds, which is necessary for filling in the above table. 
%In order to identify the $d^2$ differentials, we also need the cohomology ring structure. 
Here, note that the $\bZ_2$ cohomology ring of flag manifolds is known to be~\cite{Borel:MR51508}% see Sec.29
\begin{equation}
	\begin{array}{ccl}
		H^\ast\left(\dfrac{U(N)}{U(1)^N};\bZ_2\right)
		& \simeq &
		H^\ast\left(BU(1)^N;\bZ_2\right) \big/ I\\
		& = &
		\bZ_2[x_1, \dots, x_N] \Big/\Big(\Lambda_{\bZ}(x_1, \dots, x_N)\otimes \bZ_2\Big)
	\end{array}
	\label{eq:Z2cohomology_flag}
\end{equation}
where $\Lambda_{\bZ}(x_1,\dots, x_N)$ is a ring of symmetric functions of generators
$x_1,\dots, x_N\in H^2(BU(1)^N)$.
This implies that the differential $d^2: E^2_{4,0}\to E^2_{2,1}$
given by a dual Steenrod square composed with mod $2$ reduction~\cite{Teichner:MR1214960}
is nontrivial, and none of the $\bZ_2$'s involved survive to the $E^\infty$-page.
This shows that 
\be
\Omega^{\mathrm{spin}}_3(U(N)/U(1)^N)=0,
\ee
and thus both ordinary and discrete theta angles do not exist in the $3$d $U(N)/U(1)^N$ sigma model for $N\ge 3$.
We also find that 
\be
\widetilde{\Omega}^{\mathrm{spin}}_4(U(N)/U(1)^N)= (2\mathbb{Z})^{\oplus (N-1)} \oplus \mathbb{Z}^{\oplus \left({(N-1)(N-2)\over 2}-1\right)},
\label{eq:reduced_bordism_flag}
\ee
each factor corresponding to a Chern-Simons-like term, and there are ${N(N-1)\over 2}-1$ of them in total.\footnote{The factor $2$ in front of $\bZ$ appears in the same way as we have discussed in (\ref{eq:CPn-1_spin_4th_bordism}). Here, it means that the diagonal Chern-Simons terms are quantized with the coefficient ${1\over 4\pi}$, while the off-diagonal ones are quantized with the coefficient ${1\over 2\pi}$, which is twice of that of diagonal ones. } In this formula, we have already neglected the purely gravitational term by considering the reduced bordism. 

As we can trivially rewrite $U(N)/U(1)^{N}\simeq SU(N)/U(1)^{N-1}$, we expect that there are $N-1$ types of $U(1)$ gauge fields, $(a_I)_{I=1,\ldots, N-1}$, in the gauged sigma-model description. 
This would naively allow us to write down the $(N-1)$ diagonal Chern-Simons terms and ${(N-1)(N-2)\over 2}$ off-diagonal Chern-Simons terms, 
\begin{equation}
\sum_{I=1}^{N-1} \frac{k_I}{4\pi} \int_{W_4} f_I^2 +
\sum_{I<J}\frac{k_{IJ}}{2\pi} \int_{W_4} f_I \wedge f_J,  
\label{eq:flagCScandidate}
\end{equation}
where $f_I=\diff a_I$. 
However, this overcounts the number of independent Chern-Simons couplings according to the bordism computations~\eqref{eq:reduced_bordism_flag}. 
We shall resolve this issue in the next subsection by closely looking at the constraint on $U(1)$ gauge fields. 

In passing, we note that this is again in conflict with the homotopy argument, where the homotopy exact sequence on the fibration
\begin{equation}
	U(1)^N \to U(N) \to \dfrac{U(N)}{U(1)^N}
\end{equation}
leads to
\begin{equation}
	\renewcommand{\arraystretch}{2}
	\begin{array}{ccc}
		\pi_3\left(\dfrac{U(N)}{U(1)^N}\right) & = & \bZ,\\
		\pi_4\left(\dfrac{U(N)}{U(1)^N}\right) & = & 0.\\
	\end{array}
\end{equation}
Thus, the homotopy argument would imply the existence of one continuous theta angle. 
However, the bordism suggests that the $3$d theta terms (both ordinary and discrete) do not exist. Instead, there should to be many Chern-Simons terms, which are also invisible from the homotopy group.

\subsection{Effective Lagrangian of $3$d flag-manifold sigma models}

Let us consider the form of the effective Lagrangian for $3$d flag-manifold sigma models. 
We take an approach of the gauged nonlinear sigma model, by introducing an $SU(N)$ scalar field $U$ and $U(1)$ gauge fields $a_{I=1,\ldots, N-1}$. 
We denote the $SU(N)$-valued scalar field $U$ as 
\be
U=\left[z_1, z_2, \ldots, z_N\right], 
\ee
where $z_i$ are $N$-component complex vectors with the constraint
\be
z^\dagger_i \cdot z_j=\delta_{ij}, 
\ee
and also with $\det[z_1,\ldots, z_N]=1$. We can then express $z_N$ in terms of the other scalars $\{z_I\}_{I=1,\ldots, N-1}$ as 
\be
z_{N,\alpha}=\sum_{\alpha_1,\ldots, \alpha_{N-1}}\ve_{\alpha_1\cdots \alpha_{N-1}\alpha} z^*_{1,\alpha_1}\cdots z^*_{N-1, \alpha_{N-1}}. 
\ee
Under the $U(1)$ gauge transformations
\be
a_I(x)\mapsto a_I(x)+\diff \lambda_I(x), 
\ee
the scalar fields transform as
\be
z_I\mapsto \rme^{-\im \lambda_I}z_I
\ee
and 
\be
z_N\mapsto \rme^{\im (\lambda_1+\cdots+\lambda_{N-1})} z_N. 
\ee
We can write down the $SU(N)$-symmetric kinetic term as 
\be
{1\over g^2}\sum_{i=1}^{N} |(\diff + \im a_i)z_i|^2,
\label{eq:flag_kinetic}
\ee
where we have set $a_N:=-(a_1+\cdots+a_{N-1})$ to emphasize the $\mathbb{Z}_N$ permutation symmetry of this Lagrangian. 
We can solve the equation of motion for the gauge fields to find the constraint 
\be
a_I=\im z^\dagger_I \diff z_I,
\ee
for $I=1,\ldots, N-1$. 
With this setup, we try to find the possible topological terms that can be added to this Lagrangian. 

We have observed that there are potentially ${N(N-1)\over 2}$ Chern-Simons type terms listed in~\eqref{eq:flagCScandidate} for the 3d sigma model on the flag manifold.
Meanwhile, due to 
\be
E^2_{4,0}=H_4(SU(N)/U(1)^{N-1})=\bZ^{\oplus \left({N(N-1)\over 2}-1\right)},
\ee
the $E^2$ page of the AHSS~\eqref{eq:E2flag} implies the existence of ${N(N-1)\over 2}-1$ of them. 
This suggests that one specific linear combination of Chern-Simons terms in~\eqref{eq:flagCScandidate} somehow becomes trivial, which enables us to identify ${N(N-1)\over 2}-1$ independent Chern-Simons terms modded out by such a relation.
In Appendix~\ref{app:CSconstraints}, we actually find that the $U(1)$ gauge fields are subject to the relation given by
\be
\sum_{I=1}^{N-1}f_I^2+\sum_{I<J}f_I f_J=
-{1\over 2}\diff\left(\sum_{i\not=j<k} a_{ij}a_{jk}a_{ki}\right), 
\label{eq:flag_relation}
\ee
where we have introduced 
\be
\label{aij}
a_{ij}=z^\dagger_i \diff z_j
\ee
for $i\not =j$. Under the $U(1)$ gauge transformations, they behave as 
\bea
a_{ij}\mapsto \rme^{+\im \lambda_i}z^\dagger_i \diff (\rme^{-\im \lambda_j}z_j)=\rme^{\im(\lambda_i-\lambda_j)}\left(z^\dagger_i \diff z_j+(-\im)(z^\dagger_i \cdot z_j) \diff \lambda_j\right)=\rme^{\im (\lambda_i-\lambda_j)}a_{ij}, 
\eea
where we have used $z^\dagger_j\cdot z_i=0$ for $i\not = j$. 
Therefore, although $a_{ij}$ are not $U(1)$ gauge invariant, they transform covariantly. 
Accordingly, following combinations
\be
a_{ij}\wedge a_{jk}\wedge a_{ki}, 
\ee
for $i\not=j\not=k$, are gauge invariant\footnote{$a_{ij}a_{ji}$ is also gauge invariant, but it is not a Lorentz scalar in $3$d as it is a $2$-form. We can construct Lorentz scalars by considering $(a_{ij}a_{ji})\wedge \star (a_{k\ell}a_{\ell k})$ or $a_{ij}a_{ji}\wedge \star (\diff a_k)$, but they are  higher-derivative corrections. We note that, in $2$d, $a_{ij}a_{ji}$ is dual to the Lorentz scalar, and it indeed appears in the effective Lagrangian~\cite{Lajko:2017wif, Tanizaki:2018xto}. Therefore, we may be able to regard this new term, $a_{ij}a_{jk}a_{ki}$, as its $3$d analogue. }, 
which define globally well-defined $3$-forms. 
This shows that the right-hand-side of (\ref{eq:flag_relation}) is an exact $4$-form, and we have reproduced the result of the cohomology group by an explicit computation. 
We note that 
\be
(a_{ij}a_{jk}a_{ki})^*=a_{ji}a_{ik}a_{kj}, 
\ee
so the right-hand-side of (\ref{eq:flag_relation}) is real-valued as it should be. 

According to this relation (\ref{eq:flag_relation}), the following combination involving Chern-Simons terms
\be
\sum_{I=1}^{N-1}\frac{1}{2\pi}\int_{W_4}f_I\wedge f_I + \sum_{I<J}\frac{1}{2\pi}\int_{W_4}f_I\wedge f_J -  \sum_{i\not=j<k}\frac{1}{4\pi}\int_{M_3} a_{ij}a_{jk}a_{ki}
\ee
potentially provides a bordism invariant of 3d spin manifolds equipped with a map to $U(N)/U(1)^N$, as in the case of the Hopf term in the $\mathbb{C}P^1$ model.
Here, $M_3$ is a 3d spacetime manifold where $\partial W_4=M_3$.
However, since the spin bordism $\Omega^{\spin}_{3}(U(N)/U(1)^N)=0$,
we can take a $4$d extension as a flag-manifold field, not as $U(1)$ gauge fields. 
Hence, the above candidate term turns out to be trivial
\be
\frac{\im}{2\pi}\int_{W_4} \left(\sum_{I=1}^{N-1} f_I\wedge f_I + \sum_{I<J}f_I\wedge f_J -  {1\over 2}\sum_{i\not=j<k} \diff(a_{ij}a_{jk}a_{ki})\right)=0,
\label{eq:CSconstraintsonflags}
\ee
and we can simply eliminate one of the off-diagonal Chern-Simons term by using (\ref{eq:flag_relation}).\footnote{The relation (\ref{eq:flag_relation}) has been used to construct the Hopf invariant for the $SU(3)/U(1)^2$ flag manifold~\cite{Kisielowski:2013ina} (see also \cite{Amari:2018gbq}). 
In order to detect $\pi_3(U(N)/U(1)^N)=\mathbb{Z}$, as a special feature of the sphere $S^3$, we can take the globally defined $U(1)$ gauge fields, $a_1,\ldots, a_{N-1}$ on $S^3$, and this fact plays a crucial role to have the Hopf invariant in \cite{Kisielowski:2013ina}. 
Discussion in this paragraph has shown that the corresponding $\theta$ terms do not exist for the $3$d sigma model including discrete ones, when we adopt the bordism classification of topological terms in order to define QFTs on general $3$-manifolds.}

In the following, let us explicitly work on the case $N=3$, i.e. on the $3$d $SU(3)/U(1)^2$ flag-manifold sigma model. 
Topological terms of the effective Lagrangian can be written as 
\be
{\im k_1\over 4\pi}\int_{W_4}\diff \tilde{a}_1\wedge \diff \tilde{a}_1
+{\im k_2\over 4\pi}\int_{W_4}\diff \tilde{a}_2\wedge \diff \tilde{a}_2
+\im {h\over 2\pi}\int_{M_3}a_{12}a_{23}a_{31}. 
\label{eq:flag_topological}
\ee
Here, we note that 
\be
(a_{12}a_{23}a_{31})^*=a_{12}a_{23}a_{31}-\diff (a_{12}a_{21}), 
\ee
and thus it gives a real-valued $3$-form up to a total derivative. Therefore, we do not need to introduce coupling constants separately for $a_{12}a_{23}a_{31}$ and $(a_{12}a_{23}a_{31})^*=a_{21}a_{13}a_{32}$. 
The coupling constant $h$ for $a_{12}a_{23}a_{31}$ does not need to be quantized unlike the Chern-Simons couplings $k_{1}, k_{2}$, and it does not have a periodicity unlike the theta terms, even though the term is topological. 
Even if we add the off-diagonal Chern-Simons term, ${k_{12}\over 2\pi}\diff \tilde{a}_1\wedge \diff \tilde{a}_2$, its effect can be taken into account with the above Lagrangian by rewriting the couplings $k'_1=k_1-2 k_{12}$, $k'_2=k_2-2k_{12}$, and $h'=h+k_{12}$. 

Let us give a quick observation on how we could introduce the Chern-Simons coupling in view of symmetry. 
Our kinetic term (\ref{eq:flag_kinetic}) is constructed so that it has $\mathbb{Z}_3$ permutation symmetry, $z_i\mapsto z_{i+1 \bmod 3}$, which is often a consequence of the underlying lattice symmetry of quantum spin systems.  
However, the Chern-Simons coupling cannot be introduced in a $\mathbb{Z}_3$-symmetric way. 
Indeed, the only $\mathbb{Z}_3$ symmetric combination of the diagonal Chern-Simons term is
\be
f_1^2+f_2^2+f_3^2=2(f_1^2+f_2^2+f_1 f_2), 
\ee
since $f_3=-(f_1+f_2)$,
and because of the relation (\ref{eq:flag_relation}), this is a trivial element of the cohomology. 
This would suggest that, if we try to construct a $SU(3)/U(1)^2$ model with nontrivial Chern-Simons terms from the $SU(3)$ Heisenberg-like model on triangular lattices, we need to break the full wallpaper group, $p6m$, to some smaller subgroup explicitly. 

Lastly, we consider the properties of skyrmions for the $SU(3)/U(1)^2$ sigma model. 
As there are two independent $U(1)$ gauge fields, there are two different skyrmions charges defined by 
$Q_I={1\over 2\pi}\int_{M_2}\diff a_I$ for $I=1,2$. 
Let us introduce a skyrmion for $Q_1$ and discuss its statistics as we have done in Sec.~\ref{sec:CPN}. 
Such a skyrmion can be constructed as 
\be
U(\theta,\phi)=[z_1,z_2,z_3]
=\begin{pmatrix}
\cos{\theta\over 2} & 0 & \rme^{-\im\phi}\sin{\theta\over 2}\\
\rme^{\im\phi}\sin{\theta\over 2} & 0 & -\cos{\theta\over 2}\\
0 & 1 & 0
\end{pmatrix}. 
\ee
We can further construct a $4$d extension of the adiabatically $2\pi$-rotated skyrmions by multiplying an $SU(3)$ matrix from the left, exactly as we did in (\ref{eq:2pi_rotation_cpn-1}). 
One can readily check that the skyrmion becomes a fermion when $k_1$ is odd.
Note that the non-quantized topological term, $a_{12}a_{23}a_{31}$, does not affect the result since it vanishes identically for this configuration on $S^2\times S^1$. 

% \subsection{Statistics of skyrmions}

% Let us do some explicit computation for the $N=3$ case.
% As in Sec.~\ref{sec:CP1}, $U(3)/U(1)^3$ flag sigma model can be realized as 
% a theory with $SU(3)$ scalar field $U$ and two $U(1)$ gauge fields $a_1, a_2$.

%\newpage

%%%%%%%%%%   ACKNOWLEDGMENTS   %%%%%%%%%%

\acknowledgments
The authors thank Shuhei Ohyama and Mayuko Yamashita for valuable discussions,
and also Yuji Tachikawa for comments on the manuscript.
R.~K. and Y.~L. are grateful to the hospitality of YITP during their stays. 
The authors thank the YITP workshop YITP-W-21-11 on ``Theoretical studies of topological phases of matter'', which was useful to complete this work.
The work of R.~K. was supported by Japan Society for the Promotion of Science (JSPS) through Grant No.~19J20801. 
The work of Y.~L. was partially supported by JSPS Research Fellowship for Young Scientists.
The work of Y.~T. was partially supported by JSPS KAKENHI Grant-in-Aid for Research Activity Start-up, 20K22350.
The work of K.~S. was supported by PRESTO, JST (Grant No. JPMJPR18L4) and CREST, JST (Grant No. JPMJCR19T2).

\newpage
%---------------------------------------------------------------------------------------

\appendix

\section{Derivation of constraint relations for the flag sigma model }
\label{app:CSconstraints}

In this appendix, we prove the relation for the flag-manifold sigma model, 
\be
\sum_{i=1}^{N}f_i^2=-\diff\left(\sum_{i\not=j<k} a_{ij}a_{jk}a_{ki}\right), 
\ee
where $\sum_{i\not=j<k}$ means the summation over all $1\le i,j,k\le N$ satisfying $i\not=j$ and both $i,j <k$. 
Combined with the relation $f_{N}=-f_1-\cdots -f_{N-1}$, we find that 
\be
\sum_{I=1}^{N-1}f_I^2+\sum_{I<J}f_I f_J=
-{1\over 2}\diff\left(\sum_{i\not=j<k} a_{ij}a_{jk}a_{ki}\right), 
\ee
where capital $I,J$ run over $1,\ldots, N-1$, while $i,j,k$ run over $1,\ldots, N$. 

Let $U$ be a $U(N)$-valued scalar field and define the Maurer-Cartan form as 
\be
A=U^\dagger \diff U. 
\ee
Let $P_i$ be the projector to $i$-th component and denote
\be
A_i=P_i A P_i
\ee
and\footnote{Since $A_i$ is a $U(1)$ gauge field, $A_i^2$-term in the definition of $F_i$ is not necessary, as it automatically vanishes. However, the proof in this appendix can be generalized to the case where $A_i$ is a $U(n)$ gauge field, which corresponds to more general flag manifolds, $U(n_1+\ldots+n_N)/(U(n_1)\times \cdots\times U(n_N))$. Therefore, in the proof, we do not use any special properties for $U(1)$ gauge fields, and keep the $A_i^2$-term in the field strength.}
\be
F_i=\diff A_i +A_i^2. 
\ee
We note that, in this appendix, the gauge fields, $A$ and $A_i$, are taken to be anti-Hermitian, and thus there is an extra fator of $\im$ to make the connection with the main text of the paper:
\be
a_i=\tr(\im A_i),\quad f_i=\tr(\im F_i). 
\ee
Let us prove the following relations, 
\bea
\sum_{i}\tr(F_i)&=&0, 
\label{eq:relation_deg2}\\
\sum_{i}\tr(F_i^2)&=&\diff\left\{\sum_{i\not=j<k}\tr(P_i A P_j A P_k A)\right\} 
\label{eq:relation_deg4}. 
\eea
The first relation (\ref{eq:relation_deg2}) can be shown very easily. Since $\diff A=-A^2$, we find that 
\be
F_i=P_i(-A^2)P_i+P_i A P_i AP_i=-P_i A\sum_{j\not =i}P_j A P_i.  
\ee
Therefore, 
\bea
\sum_i \tr(F_i)&=&
-\sum_i\sum_{j\not=i} \tr(P_i A P_j A)=0. 
\eea
Here, we use the cyclic property of the trace, which gives $\tr(P_i A P_j A)=-\tr(P_j A P_i A)$.

Let us show the second relation (\ref{eq:relation_deg4}).
\bea
\sum_{i=1}^{N} \tr(F_i^2)&=&\sum_{i=1}^{N}\sum_{j,k\not=i}\tr(P_i A P_j A P_i A P_k A)\nonumber\\
&=&\sum_{I=1}^{N-1}\tr(P_I A (\bm{1}-P_I)A P_I A (\bm{1}-P_I)A) \nonumber\\
&&+\sum_{J,K} \tr\left(\left(\bm{1}-\sum_L P_L\right) A P_J A \left(\bm{1}-\sum_M P_M\right) A P_K A\right). 
\label{eq:totyushiki_constraint}
\eea
In order to obtain the last expression, we replace the summations $i,j,k$ over $1,\ldots, N$ into the summations $I,J,K$ over $1,\ldots, N-1$ and $N$ is treated separately, and we eliminate $P_N$ by substituting $P_N=\bm{1}-P_1-\cdots-P_{N-1}$. 
By cyclic property, $\tr((P_I A)^4)=0$ and $\sum_{J,K,L,M}\tr(P_J A P_K A P_L A P_M A)=0$. We therefore find 
\bea
\sum_i \tr(F_i^2)&=&
\sum_I\left\{\tr(P_I \diff A P_I \diff A)+2 \tr((P_I A)^2 P_I \diff A)\right\}\nonumber\\
&&-\sum_{J,K}\tr(P_J \diff A P_K \diff A)+2 \sum_{J,K,L}\tr(P_J A P_K A P_L \diff A)\nonumber\\
&=& \diff \left\{\sum_{J\not =K }\tr(P_J A P_K A^2)+{2\over 3}\left(\sum_I\tr((P_I A)^3)-\sum_{J,K,L}\tr(P_J A P_K A P_L A)\right)\right\}\nonumber\\
&=&\diff \left\{{1\over 3}\sum_{J\not= K\not=L}\tr(P_J A P_K A P_L A)+\sum_{J\not=K}\tr(P_J A P_K A P_N A)\right\}\nonumber\\
&=&\diff\left\{\sum_{i\not=j<k}\tr(P_i A P_j A P_k A)\right\}.
\eea
The first equation on the right-hand-side is obtained by replacing $A^2=-\diff A$ in (\ref{eq:totyushiki_constraint}) with the above note. 
As it turns out to be a total derivative, we extract the exterior derivative to find the second equality, and replace $\diff A$ by $-A^2$ inside the overall differential.  
In order to go from the second line to the third line on the right-hand-side, we have replaced $A^2=A(\sum_L P_L+P_N)A$ and then consider the combinatorics for the domain of summations. 
To obtain the last line from the third one, we note that $\tr(P_J A P_K A P_L A)$ is cyclically symmetric in $J,K,L$, so we can replace $\sum_{J\not=K\not=L}=3\sum_{J\not=K<L}$ in the first term. 
Then, we can combine the first and second terms in the third line by reintroducing the dummy indices $i,j,k$ running over $1,\ldots,N$, which gives the last line, and this completes the proof.

\section{Leray-Serre spectral sequence for cohomologies of sigma models}\label{app:LSSS}

Leray-Serre spectral sequence (LSSS) provides us a systematic way to find cohomologies from the fibration, 
\be
F\to E\to B. 
\ee
When $\pi_1(B)$ acts trivially on the cohomology of the fiber $H^\ast(F)$, LSSS gives
\be
H^p(B; H^q(F)) \Longrightarrow H^{p+q}(E),  
\label{eq:LSSS1}
\ee
that is, the $E_2$ page can be written down by knowing the cohomologies of the base space $B$ and the fiber $F$, and it converges to the cohomology of the total space $E$. 
LSSS is a first-quadrant spectral sequence, so if $B$ or $F$ is finite dimensional, then LSSS collapses at some finite page. 
There are two typical ways to use LSSS. The first one is to obtain $H^*(E)$ when we already know about $H^*(B)$ and $H^*(F)$. 
The another one is the ``opposite'' way:
knowing $H^*(E)$ and one of cohomologies of $B$ or $F$, we constrain the possible cohomologies of the last one. 

In the context of the sigma model, we are interested in the cohomology of the target space $X$. In many cases, the target space is given by the homogeneous Riemannian space $G/H$, because this appears when a global symmetry $G$ is spontaneously broken to its subgroup $H$ by some condensates. 
Then, there is a principal $H$-bundle, 
\be
H\to G\to G/H. 
\label{eq:principal_fibration}
\ee
When both $G$ and $H$ are continuous and connected Lie groups, and $G$ is moreover simply connected, then the homotopy exact sequence tells 
\be
\pi_1(G/H)\simeq 0. 
\ee
Therefore, LSSS works nicely, which gives 
\be
H^p(G/H ; H^q(H))\Longrightarrow H^{p+q}(G). 
\label{eq:LSSS2}
\ee

Since (\ref{eq:principal_fibration}) gives a principal bundle, which is not a general fibration, we have another fibration out of (\ref{eq:principal_fibration}); 
\be
G\to G/H \to \mathrm{B} H, 
\ee
where $\mathrm{B}H$ is the classifying space of $H$. 
When $H$ is connected, $\pi_1(\mathrm{B}H)\simeq \pi_0(H)=0$, and thus LSSS again works nicely. 
We obtain 
\be
H^p (\mathrm{B}H; H^q (G)) \Longrightarrow H^{p+q}(G/H). 
\label{eq:LSSS3}
\ee
Therefore, this spectral sequence makes a relationship between topological terms of $H$ gauge fields and topological terms of $G/H$-valued sigma models. 

In the following of this appendix, we first study the cohomology of $\mathbb{C}P^{N-1}$ as an exercise, and then compute and reinterpret the cohomology of $SU(N)/U(1)^{N-1}$. 

\subsection{Cohomology of $\mathbb{C}P^{N-1}$}\label{sec:LSSS_CPn-1}

Let us apply the LSSS to the fibration (\ref{eq:CPn-1_fibration}), which gives
\be
H^p(\mathbb{C}P^{N-1}; H^q(U(1)))\Longrightarrow H^{p+q}(S^{2N-1}). 
\ee
We note that $H^*(S^n)=\mathbb{Z}$ for $*=0,n$ and others vanish. Therefore, the cohomologies of the total space and the fiber is known, and thus we can constrain $H^*(\mathbb{C}P^{N-1})$ from consistency. 
Let us explicitly work on the case $N=3$, i.e. for the case of $\mathbb{C}P^2$.
The $E_2$ page is given by 
\be
\begin{sseqpage}[cohomological Serre grading, grid = chess , classes = {draw = none}, 
	yscale = 0.65, xrange = {0}{4}, 
	 x label = {$H^p(\mathbb{C}P^2)$}, y label = {$H^q(S^1)$}
	]
\foreach \x in {0,2,4}{
	\class["\mathbb{Z}"](\x,0)
	\class["\mathbb{Z}"](\x,1)
}
\d2(0,1)
\d2(2,1)
\end{sseqpage}
\ee
Since $\mathbb{C}P^{N-1}$ is connected, $H^0(\mathbb{C}P^{N-1})=\mathbb{Z}$, and then we can fill the $0$th column. Since the total-space cohomology $H^*(S^5)$ does not have the nontrivial cohomology until $*=5$, $\mathbb{Z}$ in $(p,q)=(0,1)$ must be eliminated by the $d_2$ differential. 
Let $y\in H^1(S^1)$ be the generator, and we denote $d_2(y)=x$ and  $H^2(\mathbb{C}P^{N-1})\simeq \mathbb{Z}$ is generated by $x$. 
Then, $(p,q)=(2,1)$ is generated by their tensor products $xy$, which again should be eliminated by $d_2$, and the Leibniz rule gives $d_2(xy)=d_2(x)y+x d_2(y)=x^2$. Thus, $x^2$ is the generator of $H^4(\mathbb{C}P^2)\simeq \mathbb{Z}$. Since $(p,q)=(4,1)$ must not be eliminated by differentials because $H^5(S^5)=\mathbb{Z}$, we find that $d_2(x^2y)=0$, and then we obtain $H^*(\mathbb{C}P^2)\simeq\mathbb{Z}[x]/(x^3)$ as a ring isomorphism. 
This reproduces (\ref{eq:homology_CPn-1}).

We would like to identify this generator $x$ as a $U(1)$ field strength $\diff a/2\pi$ in the gauged sigma model description. 
Since the fibration (\ref{eq:CPn-1_fibration}) is principal, we can obtain another fibration out of it: 
\be
S^{2N-1}\to \mathbb{C}P^{N-1} \to \mathrm{B}U(1). 
\ee
Let us apply the LSSS to it. We note that $\pi_1(\mathrm{B}U(1))=0$, and thus the complication of local coefficients does not come up. 
To be specific, let us consider the case $N=3$, while the discussion for general $N$ is also straightforward. 
LSSS gives 
\be
H^p(\mathrm{B}U(1); H^q(S^{5})) \Longrightarrow H^{p+q} (\mathbb{C}P^{2}). 
\ee
Then, $E_2$ page looks like 
\be
\begin{sseqpage}[cohomological Serre grading, grid = chess , classes = {draw = none}, 
	yscale = 0.65, xrange = {0}{8}, 
	 x label = {$H^p(\mathrm{B}U(1))$}, y label = {$H^q(S^5)$}
	]
\foreach \x in {0,2,4,6,8, 10, 12}{
	\class["\mathbb{Z}"](\x,0)
	\class["\mathbb{Z}"](\x,5)
}
\d[blue]6(0,5)
\d[blue]6(2,5)

\structline[blue](4,5)(10,0)
\structline[blue](6,5)(12,0)
\end{sseqpage}
\ee
As $\mathbb{C}P^2$ is a $4$-manifold, $H^*(\mathbb{C}P^2)$ must vanish for $*\ge 5$. 
Just because of this trivial information, we find that the $d_6$ differentials should be identity maps, which are drawn with blue solid arrows. 
As a result, we find that, for $N\ge 3$,
\be
H^4(\mathbb{C}P^{N-1})\simeq H^4(\mathrm{B}U(1)), 
\ee
which is generated by
\be
\int {\diff a\over 2\pi}\wedge {\diff a\over 2\pi}. 
\ee
This elucidates that the topological term of our interest can be described as the Chern-Simons coupling. 

In this case, this argument may seem to be tautological, reminding the fact that $\mathrm{B}U(1)\simeq \mathbb{C}P^{\infty}$. 
This, however, turns out to be a good exercise for the flag-manifold sigma models, as we discuss in the next subsection. 

\subsection{Cohomology of $SU(N)/U(1)^{N-1}$}
\label{sec:cohomology_flag_manifold}

Next, let us compute the cohomology of the flag manifold $SU(N)/U(1)^{N-1}$ using LSSS for $U(1)^{N-1}\to SU(N)\to SU(N)/U(1)^{N-1}$: 
\be
H^p (SU(N)/U(1)^{N-1}; H^q(U(1)^{N-1})) \Longrightarrow H^{p+q} (SU(N)). 
\ee
The basic strategy is as follows: We know the cohomologies $H^*(U(1)^{N-1})$ and $H^*(SU(N))=\bigwedge[x_3,x_5,\ldots, x_{2N-1}]$, 
and then we find out $H^*(SU(N)/U(1)^{N-1})$ from consistency. 
We also note that $\mathrm{dim}(SU(N)/U(1)^{N-1})=N(N-1)$, and the degree beyond it must vanish. 

Let us take the specific example, $N=3$, in the following. The $E_2$ page looks as
\be
\begin{sseqpage}[cohomological Serre grading, grid = chess , classes = {draw = none}, yscale = 0.65, 
	 x label = {$H^p(SU(3)/U(1)^2)$}, y label = {$H^q(U(1)^2)$}
	]
\class["\mathbb{Z}"](0,0)
\class["\mathbb{Z}^{\oplus 2}"](0,1)
\class["\mathbb{Z}"](0,2)
\class["\mathbb{Z}^{\oplus 2}"](2,0)
\class["\mathbb{Z}^{\oplus 4}"](2,1)
\class["\mathbb{Z}^{\oplus 2}"](2,2)
\d2(0,1) \d2(0,2)
\class["\mathbb{Z}^{\oplus 2}"](4,0)
\class["\mathbb{Z}^{\oplus 4}"](4,1)
\class["\mathbb{Z}^{\oplus 2}"](4,2)
\d2(2,1) \d2(2,2)
\class["\mathbb{Z}"](6,0)
\class["\mathbb{Z}^{\oplus 2}"](6,1)
\class["\mathbb{Z}"](6,2)
\d2(4,1) \d2(4,2)
\end{sseqpage}
\ee
Here, we denote $\mathbb{Z}^{\oplus n}=\oplus_{i=1}^n \mathbb{Z}$. 
As a result, we find that 
\be
H^p(SU(3)/U(1)^2)=\left\{\begin{array}{cc}
\mathbb{Z} & \quad p=0,6,\\
\mathbb{Z}^{\oplus 2} & \quad p=2,4, \\ 
0 & \quad \mathrm{else}.
\end{array}\right.
\label{eq:cohomology_flag_SU(3)}
\ee
Although we did not identify the cohomology ring structure, it is consistent with (\ref{eq:Z2cohomology_flag}) as a module.\footnote{Indeed, LSSS is not necessarily useful to determine the cohomology ring structure. In particular, the cohomology ring of the total space cannot be determined from LSSS. For example, we can consider the fibration, $\mathbb{C}P^1\to U(3)/U(1)^3\to \mathbb{C}P^2$, which can be obtained from $U(2)\times U(1)\to U(3)\to \mathbb{C}P^2$ by dividing the total space and the fiber by $U(1)^3$. LSSS immediately reproduces (\ref{eq:cohomology_flag_SU(3)}) as a module, but it does not give the cohomology ring structure. } 
With this information, we can start the computation of the spin bordism using AHSS, as we did in the main text. 
Here, instead, let us take a close look at the physical meaning of these nontrivial cohomologies. 
As in the case of $\mathbb{C}P^{N-1}$ model, it is convenient to describe the $SU(N)/U(1)^{N-1}$ sigma model using the $SU(N)$-valued scalar field $U$ with $(N-1)$ $U(1)$ gauge fields, $a_1,\ldots, a_{N-1}$. 
Then, the natural question is whether the nontrivial cohomologies of $SU(N)/U(1)^{N-1}$ are related to the nontrivial topological terms of these $U(1)$ gauge fields. 

The answer turns out to be affirmative. To see this, we consider the following fibration, 
\be
SU(N)\to SU(N)/U(1)^{N-1} \to \mathrm{B} U(1)^{N-1}.
\ee
We apply LSSS to this fibration. Let us set $N=3$ again for simplicity, then the $E_2$ page is given by 
\be
\begin{sseqpage}[cohomological Serre grading, grid = chess , classes = {draw = none}, 
	yscale = 0.65, xscale = 1, xrange = {0}{8}, 
	 x label = {$H^p(\mathrm{B}U(1)^2)$}, y label = {$H^q(SU(3))$}
	]
\foreach \y in {0,3,5,8}{
	\class["\mathbb{Z}"](0,\y)
}
\foreach \y in {0,3,5,8}{
	\class["\mathbb{Z}^{\oplus 2}"](2,\y)
}
\foreach \y in {0,3,5,8}{
	\class["\mathbb{Z}^{\oplus 3}"](4,\y)
}
\foreach \y in {0,3,5,8}{
	\class["\mathbb{Z}^{\oplus 4}"](6,\y)
}
\foreach \y in {0,3,5,8}{
	\class["\mathbb{Z}^{\oplus 5}"](8,\y)
}
\foreach \y in {0,3,5,8}{
	\class(10,\y)
	\class(12,\y)
}

\foreach \x in {0, 2, 4} \foreach \y in {3, 8}{
	\d[blue]4(\x,\y)
}

\structline[blue](6,3)(10,0) \structline[blue](6,8)(10,5)
\structline[blue](8,3)(12,0)

\draw[differential style, red, densely dashed](0,5) -- (6,0); 
\draw[differential style, red, densely dashed](2,5)--(8,0); 
\structline[red, densely dashed](4,5)(10,0) \structline[red, densely dashed](6,5)(12,0)
\end{sseqpage}
\ee
Here, the $d_4$ differentials are drawn with blue solid arrows, and the $d_6$ differentials are drawn with the red dashed arrows. 
The spectral sequence clashes after the $E_6$ page. 
We then find that only the $q=0$ components with $H^p(\mathrm{B}U(1)^2)$ for $p=0,2,4,6$ can survive in the $E^{\infty}$ page, and thus the nontrivial cohomology $H^* (SU(3)/U(1)^2)$ has a well-defined origin in  $H^p(\mathrm{B}U(1)^2)$. 

For general $N$, we get
\be
H^2(SU(N)/U(1)^{N-1})=\mathbb{Z}^{\oplus (N-1)},\quad 
H^4(SU(N)/U(1)^{N-1})=\mathbb{Z}^{\oplus \left({N(N-1)\over 2}-1\right)}. 
\ee
We note that $H^2(SU(N)/U(1)^{N-1})\simeq H^2(\mathrm{B}U(1)^{N-1})\simeq \mathbb{Z}^{\oplus (N-1)}$. As we have seen in the case $N=3$, however, 
\bea
H^4(SU(N)/U(1)^{N-1})&\simeq& H^4(\mathrm{B}U(1)^{N-1})/d_4^{0,3}(H^3(SU(N))) \nonumber\\
&\simeq& \mathbb{Z}^{\oplus {N(N-1)\over 2}}/d_4^{0,3}(\mathbb{Z}) \nonumber\\
&\simeq& \mathbb{Z}^{\oplus \left({N(N-1)\over 2}-1\right)}. 
\eea
Thus, we can express these topological terms as $f_i\wedge f_j$, 
but there is a constraint compared with the $U(1)^{N-1}$ gauge theory. 
We have computed the explicit relation among them in Appendix~\ref{app:CSconstraints}.

\section{
\label{sec:fermion_sigma}
Realization of topological terms by a free massive fermion}

In this section, we present an explicit realization of the Hopf term \eqref{eq:definition_Hopf} as a partition function of a massive free fermion. 
Let us consider a $(2+1)$d Euclidean relativistic Dirac fermion coupled with spin fields $\bm{n}$ which take values on sphere $S^2$, 
\begin{align}
{\cal L} = \im \bar \psi (\slashed{D} + m \bm{n} \cdot \bm{\s}) \psi, \quad \bm{n}^2=1. 
\label{eq:3d_dirac}
\end{align}
Here, $\slashed{D} = \g_\mu (\p_\mu-\im A_\mu)$ is a covariant derivative with $A_\mu$ a spin$^c$ gauge field, $\g_\mu$'s are the 2 by 2 gamma matrices, and $\bm{\s}$ are Pauli matrices for the internal spin space. 
It was shown that the large-scale dynamics of the spin fields $\bm{n}$ compared to the inverse of the mass, $1/m>0$, is described by the nonlinear sigma model with the Hopf term~\cite{Abanov:1999qz}, 
\begin{align}
\exp\left({-S_{\rm eff}}\right) = \int D \bar \psi D \psi\, \exp\left({- \int d^3 x {\cal L}}\right), 
\end{align}
where\footnote{
Strictly speaking, the imaginary part of \eqref{eq:spinc-hopf} is defined as
\begin{align}
\frac{i}{4\pi} \int_{W_4} (\diff \tilde a + 2 \diff \tilde A) \wedge \diff \tilde a
\end{align}
with $\tilde a$ and $\tilde A$ extensions of $U(1)$ and spin$^c$ fields, respectively, to a spin$^c$ bounding manifold $W_4$.}
\begin{align}
S_{\rm eff} = \frac{m}{8 \pi} \int d^3 x (\p_\mu \bm{n})^2 + \frac{\im}{4\pi} \int a \diff a + \frac{\im}{2\pi} \int A \diff a. 
\label{eq:spinc-hopf}
\end{align}
The imaginary part of the effective action $S_{\rm eff}$ is the Chern-Simons term corresponding to the spin$^c$ bordism group $\tilde \Omega^{\rm spin^c}_4(\C P^1) = \Z$, and it describes two topological responses: a skyrmion of the spin fields $\bm{n}$ has a unit $U(1)$ charge, and the adiabatic $2\pi$ rotation of the skyrmion texture gives rise to the Berry phase $(-1)$ on the ground state wave function. 
By restricting the background spin$^c$ fields $A$ to a spin one, the third term of \eqref{eq:spinc-hopf} vanishes and only the $\Z_2$-quantized Hopf term of \eqref{eq:spinc-hopf} remains, which is nothing but the Hopf term \eqref{eq:definition_Hopf} with $k=1$. 

A corresponding lattice model of \eqref{eq:spinc-hopf} is the following imaginary-time-dependent Hamiltonian of fermions on a closed two-dimensional square lattice \cite{Qi:2008ew}
\begin{align}
\hat H(\tau)
&= \sum_{\bx \in \Z^2} \Bigg[  
\left(\hat \psi^\dag_{\bx+\hat x} \frac{-\s_z\tau_z+\im\tau_x}{2} \hat \psi_\bx \nonumber +
\hat \psi^\dag_{\bx+\hat y} \frac{-\s_z\tau_z+\im\tau_y}{2} \hat \psi_\bx 
+ {\rm h.c.}\right)\\
& \qquad \qquad + \hat \psi^\dag_\bx 
(m \bm{n}(\bx,\tau) \cdot \bm{\s} \tau_z + \lambda \s_z \tau_z) \hat \psi_\bx
\Bigg], 
\label{eq:lattice_skyrmion}
\end{align}
where $\hat \psi^\dag_\bx/\hat \psi_\bx$ are complex fermion creation/annihilation operators, $\s_{\mu}, \tau_\mu$ are Pauli matrices, $\bm{n}(\bm{x},\tau)$ are spin fields with $\bm{n}^2=1$, and $m>0$ and $\lambda$ are parameters controlling whether the skyrmion becomes a fermion or not. 
For a uniform configuration of the spin fields $\bm{n}(\bm{x},\tau) \equiv \underline{\bm{n}}$, this Hamiltonian can be written as $\hat H= \sum_\bk \psi^\dag_\bk {\cal H}(\bk,\underline{\bm{n}}) \psi_\bk$ with a one-particle Hamiltonian 
\begin{align}
{\cal H}(\bk,\underline{\bm{n}}) = \sin k_x \tau_x + \sin k_y \tau_y + (m \underline{\bm{n}}\cdot \bm{\s}+(\lambda-\cos k_x - \cos k_y)\s_z)\tau_z 
\end{align}
in the Bloch-momentum space. 
The energy eigenvalues of ${\cal H}(\bk,\underline{\bm{n}})$ are given by 
\begin{align}
\varepsilon(\bk)^2
= \sin^2 k_x + \sin^2 k_y + m^2 \underline{n}_1^2+m^2 \underline{n}_2^2+(m \underline{n}_3+\lambda-\cos k_x-\cos k_y)^2, 
\end{align}
which has a finite energy gap unless $m=|\lambda|, |\lambda+2|, |\lambda-2|$ is satisfied. 
For $\lambda =2$ the low-energy Hamiltonian around the Gamma point $\bk \sim \bm{0}$ reduces to the Dirac model \eqref{eq:3d_dirac}, thus it is expected that the ground state with a skyrmion indicates desired topological responses for an appropriate parameter region.
Hereafter, we set $m=1$ and focus on $\lambda = 0, \pm 2, \pm 4$ as representatives of different invertible phases.

A skyrmion texture in a finite square lattice with sites $\{1,\dots, L_x\} \times \{1,\dots, L_y\}$ in the periodic boundary conditions is introduced, for example, by setting the unit vector $\bm{n}$ to 
\begin{align}
&n_1(x,y,\tau=0) = \sin\left(\frac{2\pi x}{L_x}-\pi\right)\cos\left(\frac{\pi y}{L_y}-\frac{\pi}{2}\right), \\
&n_2(x,y,\tau=0) = \sin\left(\frac{2\pi x}{L_x}-\pi\right)\cos^2\left(\frac{\pi x}{L_y}-\frac{\pi}{2}\right), \\
&n_3(x,y,\tau=0) = 1-2\cos^2\left(\frac{\pi x}{L_x}-\frac{\pi}{2}\right)\cos^2\left(\frac{\pi y}{L_y}-\frac{\pi}{2}\right). 
\end{align}
This construction is based on the smash product $S^1 \wedge S^1 \cong S^2$ \cite{Higashikawa:2018qsh}.
We numerically find that the number of electrons of the ground state $\ket{\Omega}$ (equivalently, the number of negative eigenvalues of one-particle Hamiltonian of \eqref{eq:spinc-hopf}) in the presence of a skyrmion is 
\begin{align}
\Delta N := \braket{\Omega|\hat N|\Omega} - 2L_xL_y 
=
\left\{\begin{array}{ll}
-1 & (\lambda=\pm 2),\\
2 & (\lambda=0),\\
0 & (\lambda=\pm 4), 
\end{array}\right. 
\label{eq:electron_number_skyrmion}
\end{align}
for sufficiently large $L_x$ and $L_y$. 
Here, $2L_xL_y$ is the number of electrons of the ground state without skyrmions.  
Thus, we confirm that a skyrmion behaves as an electron for $1<|\lambda|<3$. 
We also confirm that $\Delta N$ coincides with the second Chern number $-\frac{1}{8 \pi^2} \int_{T^2 \times S^2} \tr {\cal F}(\bk,\bm{n})^2$ of the semiclassical Hamiltonian ${\cal H}(\bk,\bm{n})$ over the Bloch-momentum space $T^2$ and the spin space $S^2$ with the identity map $\bm{n}: S^2 \to S^2$. 

Next, we evaluate the imaginary-time evolution of the Hamiltonian with the $2\pi$ rotation of skyrmion, which can be introduced by hand as 
\begin{align}
&\tilde n_1(x,y,\tau)+ \im \tilde n_2(x,y,\tau) = \rme^{2\pi \im\tau/T} (n_1(x,y,\tau=0)+\im n_2(x,y,\tau=0)), \nonumber \\
&\tilde n_3(x,y,\tau) = n_3(x,y,\tau=0). 
\end{align}
For the adiabatic limit $T \to \infty$, the ground state $\ket{\Omega}$ stays in the instantaneous ground state $\ket{\Omega(\tau)}$, which is defined by $\hat H(\tau) \ket{\Omega(\tau)} = E_\Omega(\tau) \ket{\Omega(\tau)}$, and acquires the Berry phase as in 
\begin{align}
P \rme^{-\int_0^T \hat H(\tau)\diff \tau} \ket{\Omega}
\sim \rme^{-\int_0^T E_{\Omega}(\tau') \diff\tau'} \rme^{-\oint \braket{\Omega(\tau)|\diff\Omega(\tau)}} \ket{\Omega}.  
\end{align}
The Berry phase $\rme^{-\oint \braket{\Omega(\tau)|\diff\Omega(\tau)}}$ can be computed as the product of fermion ground-state overlaps
\begin{align}
\braket{\Omega(\tau+\delta \tau)|\Omega(\tau)}
= \det [\Psi(\tau+\delta \tau)^\dag \Psi(\tau)]. 
\end{align}
Here, $\Psi(\tau) = (\phi_1(\tau),\dots, \phi_{N}(\tau))$ is the $4L_xL_y \times N$ matrix composed of one-particle eigenvectors $\phi_j(\tau), j=1,\dots,N$, which are eigenvectors of the one-particle Hamiltonian defined in \eqref{eq:lattice_skyrmion} having negative energies. 
We then numerically find that 
\begin{align}
\rme^{-\oint \braket{\Omega|\diff\Omega}}
= \left\{\begin{array}{ll}
-1 & (\lambda=\pm 2), \\
1 & (\lambda=0, \pm 4), 
\end{array}\right. 
\end{align}
as expected. 
The skyrmion with an odd $U(1)$ charge has an odd half-integer spin. 

It is not evident from the point of view of the lattice model that the Berry phase of the Hamiltonian is always quantized to $\pm 1$. 
We leave this issue to future studies.

% It is interesting to see more direct relationship between the fermion parity and the $2\pi$ rotation. 
% Notice that the imaginary-time dependent Hamiltonian $\hat H(\tau)$ can be written as the unitary transformation 
% \begin{align}
% \hat H(\tau)
% = 
% \hat U(\tau) \hat H \hat U(\tau)^\dag
% \end{align}
% with 
% \begin{align}
% \hat U(\tau) = e^{-i \frac{2\pi \tau}{T} \sum_\bx \psi^\dag_\bx \frac{\s_z}{2} \psi_\bx}. 
% \end{align}
% Note that although $\hat U(\tau)$ is not periodic by $T$ as $\hat U(T) = (-1)^{\hat N}$, the Hamiltonian $\hat H(\tau)$ itself is periodic. 
% In this case the instantenous ground state is written as 
% \begin{align}
% \ket{\Omega(\tau)}
% =e^{-\pi i N \frac{\tau}{T}} \hat U(\tau)\ket{\Omega}, \quad 
% N = \braket{\Omega|\hat N|\Omega}. 
% \end{align}
% The prefactor on the right-hand-side is needed for the periodicity of $\ket{\Omega(\tau)}$. 
% Then, the Berry phase is simply written as the ground state expectations 
% \begin{align}
% e^{-\oint \braket{\Omega|d\Omega}} 
% &= 
% (-1)^N \times 
% e^{-\braket{\Omega|\int_0^T U^\dag(\tau)|\frac{d U}{d\tau}(\tau)|\Omega}}\\
% &=
% (-1)^N \times 
% e^{2\pi i \braket{\Omega|\sum_\bx \psi^\dag_\bx \frac{\s_z}{2} \psi_\bx |\Omega}}. 
% \end{align}
% This, if the magnetization $M_z = \braket{\Omega|\sum_\bx \psi^\dag_\bx \frac{\s_z}{2} \psi_\bx |\Omega}$ is an integer, the Berry phase matches with the fermion parity. 
% We find that $M_z$ numerically vanishes...

%\bibliographystyle{JHEP}
\bibliographystyle{utphys}
\bibliography{./QFT,./refs}
\end{document}